\documentclass[journal]{IEEEtran}

\ifCLASSINFOpdf
\else
   \usepackage[dvips]{graphicx}
\fi
\usepackage{url}
\usepackage{citesort}

\usepackage{graphicx}
\usepackage{balance}
\usepackage{cancel}
\usepackage{booktabs}
\usepackage{multirow,multicol}
\usepackage{siunitx}
\usepackage{array,tabularx,amstext,booktabs}
\usepackage[normalem]{ulem}
\usepackage[table]{xcolor}
\usepackage{afterpage,float}

\colorlet{grfl}{gray!20}

\newcolumntype{Y}{>{\centering\arraybackslash}X}

\begin{document}

\title{Corrections to Published Values of Frequency Sampling Filter
  Transition Coefficients}

\author{C.S. Ramalingam
\thanks{The author is with the Department of Electrical Engineering,
  IIT Madras, Chennai--600036 India (email: csr@ee.iitm.ac.in).}}

\maketitle

\begin{abstract}
  Tables of optimal transition coefficients and peak sidelobe level
  (PSL, in dB) associated with frequency sampling filter (FSF) design
  were published by Rabiner et al. (Jun 1970), and reproduced, for
  example, in the book \textit{Digital Signal Processing} by Proakis
  and Manolakis (4/e, 2007).  A set of values are also given in
  Appendix~H of \textit{Understanding Digital Signal Processing} by
  Lyons (3/e, 2011), but there are significant differences between
  these two sets.  For example, for $N=16$ and $\mbox{BW}=4$, two
  different transition coefficient values have been reported, viz.,
  $0.38916626$ (Rabiner, et al.) and $0.34918551$ (Lyons).
  \textit{Neither is optimal}, for we find the optimum value to be
  $0.40474097$.  The published values of the corresponding PSLs were
  also found to be incorrect.  In this paper we give the optimal
  values of the transition coefficients and PSL values as estimated by
  our program for the lowpass and bandpass filters listed in Rabiner
  et al. and Lyons.
\end{abstract}

\begin{IEEEkeywords}
Frequency sampling, digital filter, transition coefficient
\end{IEEEkeywords}

\IEEEpeerreviewmaketitle

\section{Introduction}
\IEEEPARstart{T}{he} design of digital filters using the method of
frequency sampling was pioneered by Rabiner et al.
\cite{rabiner-gold-mcgonegal-1970,rabiner-schafer-1971,rabiner-bstj-1972,rabiner-1972}
in the early 1970s.  Its advantages are well-known
\cite{rabiner-gold-75,jackson-96} but this approach largely fell out
of favor after the advent of the Parks-McClellan algorithm
\cite{parks-mcclellan-1972,oppenheim-dtsp-3e,jackson-96}.  Lyons has
devoted thirty-five pages to frequency sampling filters (FSFs) in his
book \cite{lyons-understanding-dsp-3e}, titling Section~7.5 as
``Frequency Sampling Filters: The Lost Art'' (Appendix~G furnishes
detailed derivations and Appendix~H has extensive tables of transition
coefficients, akin to what was provided in
\cite{rabiner-gold-mcgonegal-1970}).

The archetypal lowpass filter shown in \cite[Fig.~2]{rabiner-1972} has
two transition samples $T_1$ and $T_2$ that are adjusted so as to
minimize the maximum stopband error $|T_3|$.  In this context, the
parameter BW denotes the number of consecutive frequency samples that
have unity magnitude, with the filter length $N$ being the other key
parameter.  In Table~I of \cite[p.~92]{rabiner-gold-mcgonegal-1970}
(and in succeeding tables) the values of one or more transition
samples are listed for a given value of $N$, BW, and type of filter
(lowpass, etc.).  A selection from these tables has been published in
\cite{proakis-2007}.  These are claimed to be the optimum values and,
if they indeed were, should result in FSFs with the smallest maximum
stopband erorr.  In the next section we show that, by comparing them
with the results of our optimization program (based on linear
programming), many of the published values of the transition
coefficients are \textit{not} optimum.  In
Section~\ref{sec:fsf_tables} we present our estimates of the optimal
values for the lowpass and bandpass filters given in
\cite{lyons-understanding-dsp-3e} and
\cite{rabiner-gold-mcgonegal-1970}.

\section{Non-Optimality of Published Values}
In the design of FSFs the non-trivial part of the procedure is finding
the optimal transition coefficients.  Once these values are known, one
can easily verify the stopband minimax error by finding the maximum of
the highest sidelobe (in dB) in the stopband.  This is called as the
peak sidelobe level (PSL)\footnote{Labeled as ``Atten'' in
\cite{lyons-understanding-dsp-3e} and ``minimax'' in
\cite{rabiner-gold-mcgonegal-1970}}.  Hence, if there are two competing
values for the transition coefficients, it a simple exercise to find
out the one that results in a lower PSL.

Consider $N=32$ and $\mbox{BW}=6$.  The optimal transition value as
reported in \cite[Fig.~H-1(a)]{lyons-understanding-dsp-3e} is
$0.37172559$, with the minimax error being $–45.5$.  For the same
filter, the values reported in \cite{rabiner-gold-mcgonegal-1970} are
$0.37897949$ and $-40.85183477$.  Clearly, both cannot be optimal.  In
fact, it turns out that neither is.  The optimum values were found to be
$0.39201042$ and $-42.4640$.  When the software accompanying
\cite{rorabaugh-dsp-primer} was run on this example it returned values
of $0.392011$ and $-42.464$, which are practically the same as our
estimates.

\begin{figure}[h!]
\centerline{\includegraphics[width=\columnwidth]{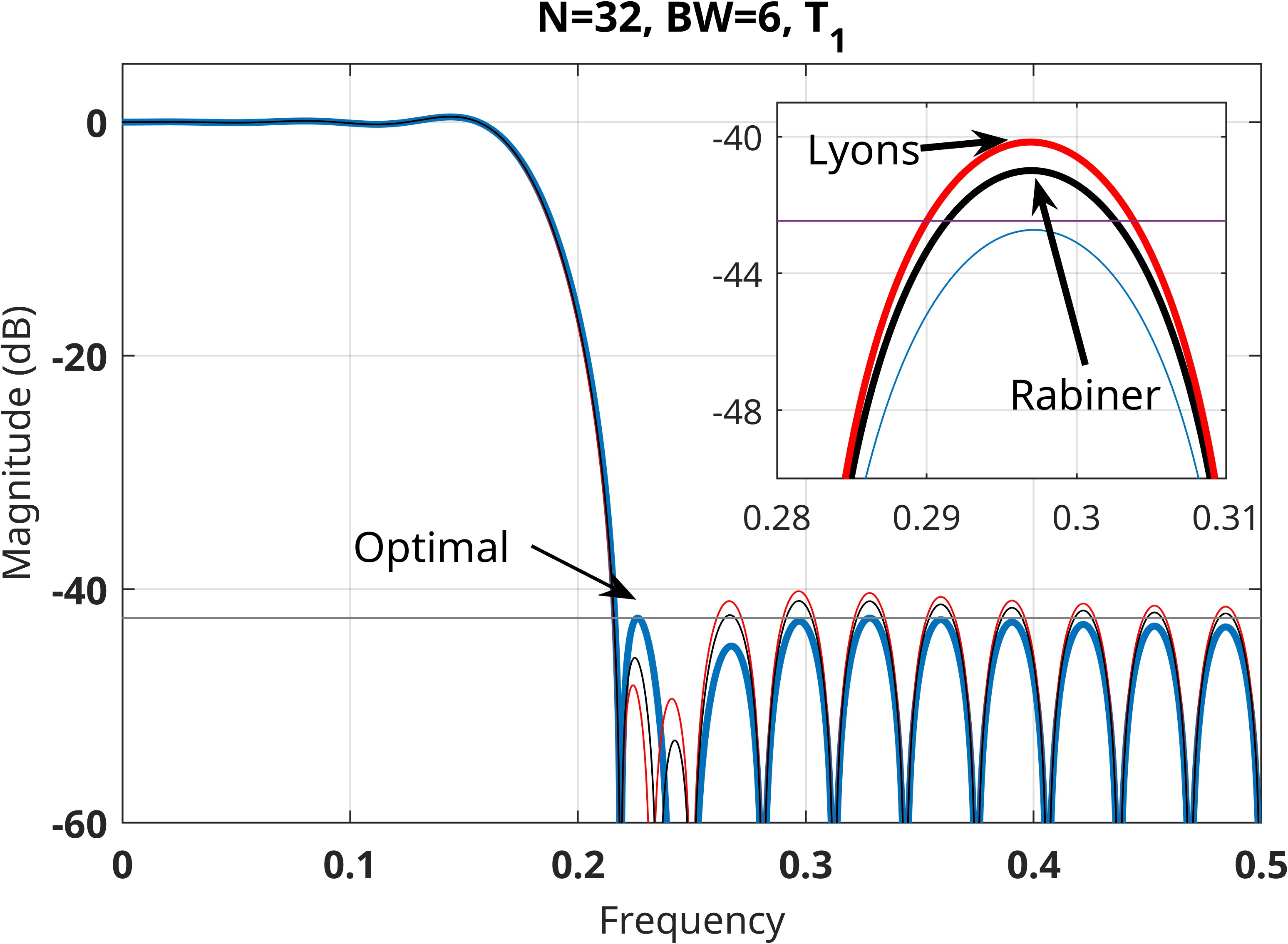}}
\caption{The values of the transition coefficients given by Rabiner,
  et al. \cite{rabiner-gold-mcgonegal-1970} and Lyons
  \cite{lyons-understanding-dsp-3e} result in PSL values of $-40.99$
  and $-40.16$, respectively.  The optimum value of $0.39201059$
  results in a PSL value of $-42.464$ (thin horizontal line).  Hence,
  the published values of both transition coefficients and PSLs are
  incorrect.}
\label{fig:nonopt_32_bw6_single}
\end{figure}
The frequency responses shown Fig.~\ref{fig:nonopt_32_bw6_single}
reveal why the published values are not optimal.  From these curves it
can be seen that the optimum coefficient's PSL is $-42.464$.  This is
also the same value returned by our optimization program, which is
satisfying.  On the other hand, for a transition coefficient of
$0.37897949$ (Rabiner, et al.), the actual PSL was found to be
$-40.9934$ instead of the published value of $-40.85183477$.
Similarly, for a transition coefficient of $0.37172559$ (Lyons), the
actual PSL was found to be only $-40.1590$ instead of the published
value of $-45.5$.  For each of these coefficient values our PSL
estimates match with those obtained using Rorabaugh's software
\cite{rorabaugh-dsp-primer}.  Hence both transition coefficients and
PSL values given in
\cite{lyons-understanding-dsp-3e,rabiner-gold-mcgonegal-1970} are not
optimal.

We also considered $N=16$, $\mbox{BW}=4$, and one transition sample
because the reported PSL value of $-49.6$ in
\cite{lyons-understanding-dsp-3e} appeared to be suspiciously low.
Our estimates of the optimum coefficient and PSL were $0.40474097$ and
$-41.6636$ (same as that obtained using Rorabaugh's software).  This is
quite different from $0.34918551$ given in Table~H-1
\cite[p. 886]{lyons-understanding-dsp-3e}, with the corresponding PSL
being only $-34.7897$.  Hence, neither the published coefficient is
optimal nor is the reported PSL corresponding this value correct.
The nearly $7\,$dB difference in this example illustrates how
significantly in error some of the published values are.

We next considered Example~13.5 \cite[p.~262]{rorabaugh-dsp-primer},
which was a $21$-tap filter with $\mbox{BW}=5$ and two transition
samples.  Our values were $T_1 = 0.57952231$ and $T_2=0.09850646$,
which is a close match to the given values, i.e., $T_1=0.579666$ and
$T_2=0.0985806$ \cite[Table~13.8, p.~264]{rorabaugh-dsp-primer}.  The
larger difference between our results and Rorabaugh's is due to the
values of the tolerance parameter used.  We could not get a closer
match by specifying a tighter tolerance because the provided
executable had convergence problems when more than one transition
coefficient was specified.  It also did not work for bandpass filters
even with one coefficient.  We did not attempt to fix bugs and
recompile this old software to make it work for all its intended cases
because non-trivial changes were required to make it compatible with
the current C++ standards.

\begin{figure}[b!]
\centerline{\includegraphics[width=\columnwidth]{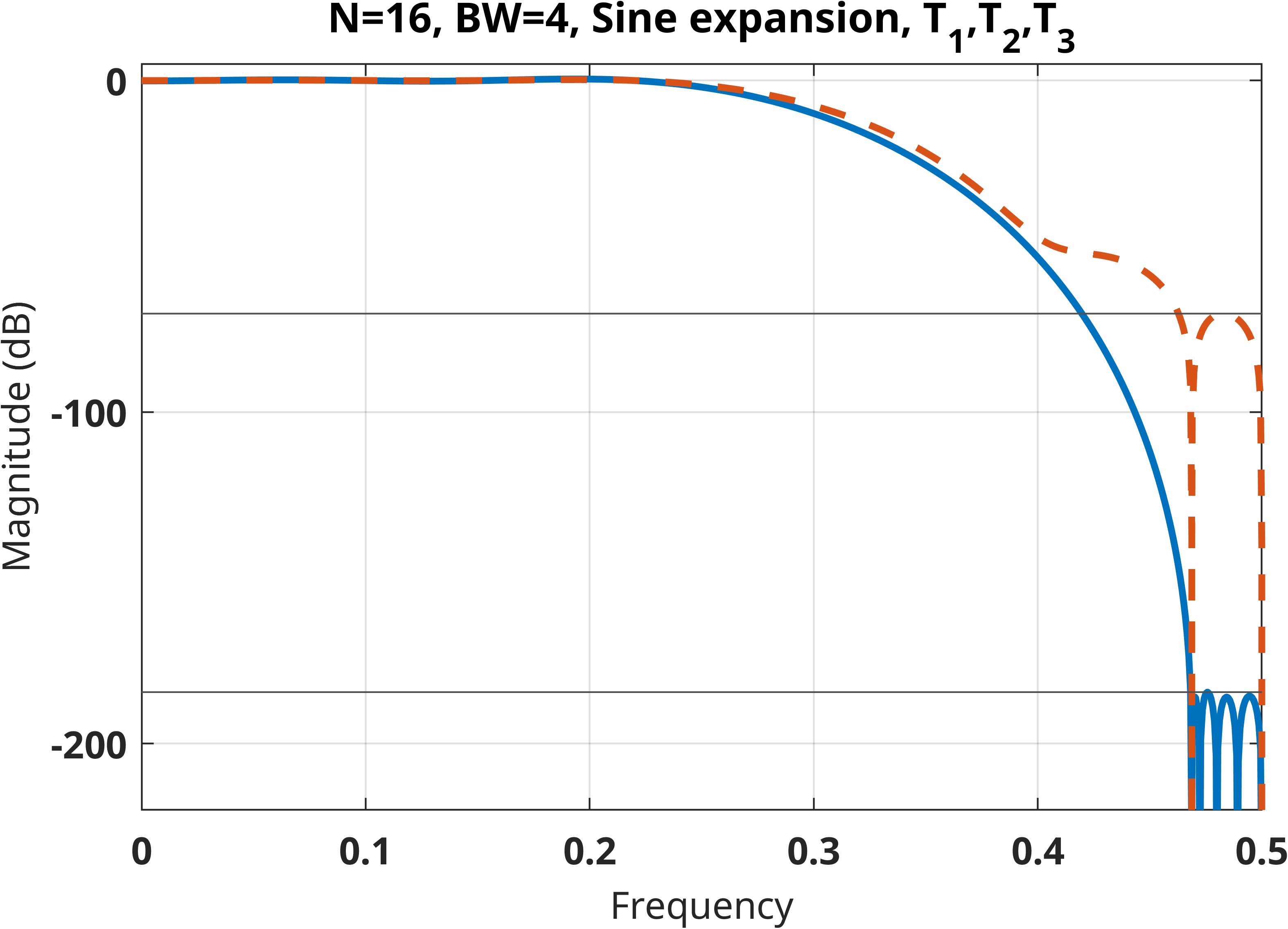}}
\caption{For an LPF with $N=16$, $\mbox{BW}=4$, three transition
  coefficients, and sine expansion, the published values are very far
  from optimal.  The reported PSL of $-129.92168427$ is wrong, with
  correct value being only $-70.3304$ (dashed curve).  The optimal
  filter's PSL value is $-184.4643$ (solid curve), which is better by
  more than $114\,$dB!}
\label{fig:n_16_bw_4_sine_exp_one_coeff}
\end{figure}
An example in which not only is the published PSL value wrong but also
its actual value being higher by \textit{more than $114\,$dB} compared
with that of the optimal filter is the $16$-tap LPF that uses the sine
expansion (labeled as Type-2 in \cite{rabiner-gold-mcgonegal-1970})
and having three transition coefficients.  For $\mbox{BW}=4$, the
reported PSL is $-129.92168427$.  The published coefficients are
$0.58824614$, $0.10690445$, and $0.00327759$, but the PSL
corresponding to these is only $-70.3304$.  The optimal coefficients
and PSL were found to be $\{0.48812533, \,0.07000579, \,0.00122350\}$
and $-184.4643$, which is better by more than $114\,$dB!  (see
Fig.~\ref{fig:n_16_bw_4_sine_exp_one_coeff}).

We tested our program on a $118$-tap bandpass filter with
$\mbox{BW}=11$ and having transition coefficients at $k=23,\,35$
\cite{rybka-bpf-2020}.  For such cases, it is usual to constrain the
coefficients in both the transition regions to be the same, which
halves the number of estimates.  Our optimization resulted in
$H[23] = H[35] = 0.38765962$ and $\mbox{PSL}=-41.2769$.  The
transition coefficient value reported in \cite{rybka-bpf-2020} is
$0.385346$, whereas its actual PSL was found to be slightly worse,
viz., $-40.9800$.

\begin{figure}[b!]
\centerline{\includegraphics[width=\columnwidth]{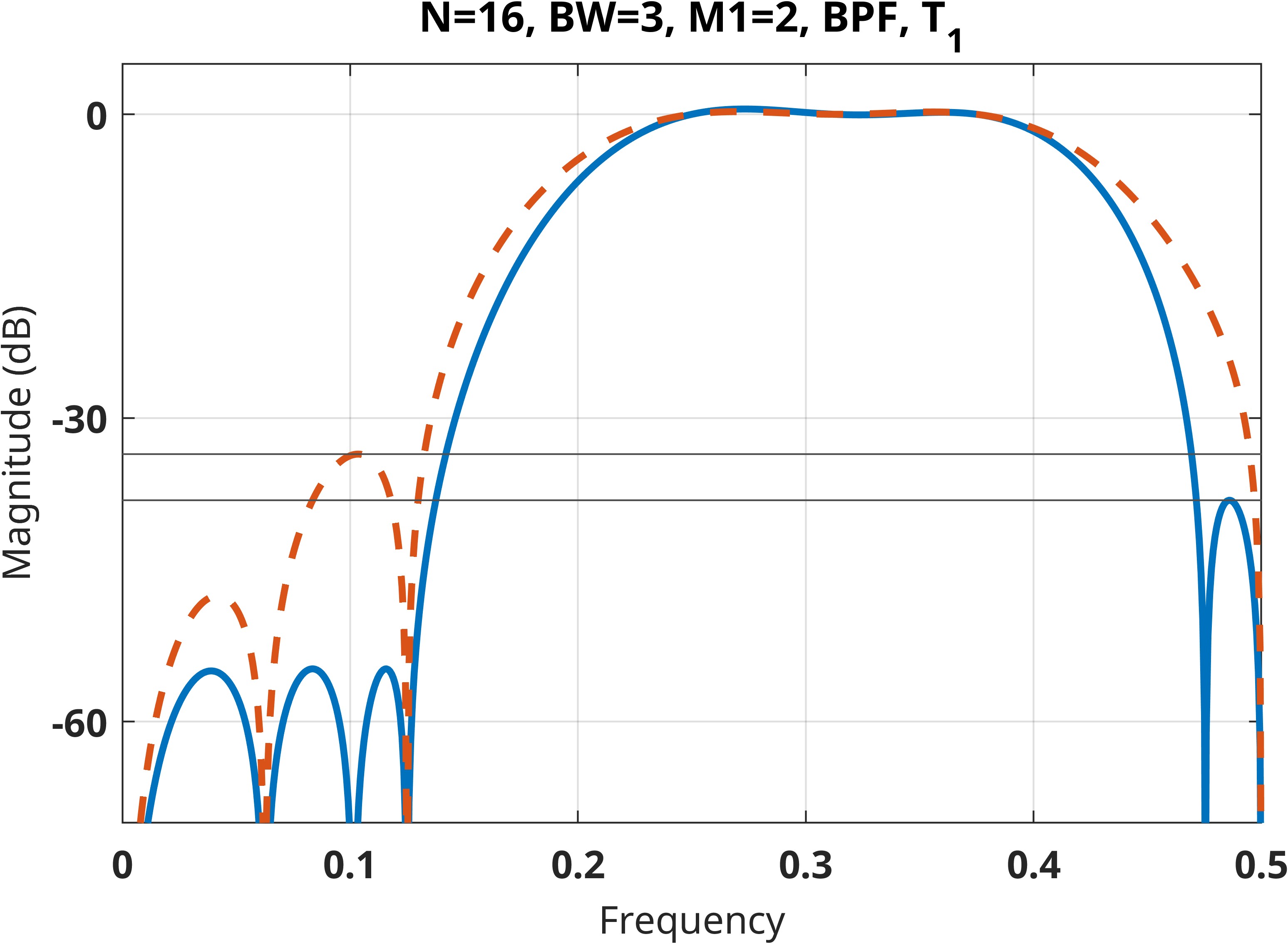}}
\caption{Frequency response of BPF with $N=16$, $\mbox{BW}=4$,
  $\mbox{M1}=2$, and $T_1= 0.31833319$ (solid curve).  Its PSL equals
  $-38.1280$.  Also shown is Rabiner et al.'s non-optimal filter
  (dashed curve) with $T_1=0.45593262$, with PSL $-33.5714$.  The
  published PSL value of $-34.175276$ is incorrect (by about
  $0.6\,$dB).}
\label{fig:n_16_bw_3_m1_2_bpf_one_trans}
\end{figure}

The very first BPF in Table~XI \cite{rabiner-gold-mcgonegal-1970} is
non-optimal whose actual PSL is nearly $5\,$dB worse.  Its parameters
are $N=16$, $\mbox{BW}=3$, $\mbox{M1}=2$, and one transition
coefficient.  The published values are $0.45593262$ and $-34.175276$,
but the actual PSL for this coefficient is marginally poorer, i.e.,
$-33.5714$.  The optimal coefficient and PSL values are $0.31833319$
and $-38.1280$ (see Fig.~\ref{fig:n_16_bw_3_m1_2_bpf_one_trans}).

An example BPF whose published PSL is incorrect by nearly $25\,$dB is
the $32$-tap filter with $\mbox{BW}=2$ and $M1=5$.  The reported
coefficients $\{0.00422363,\,0.08800332,\,0.46619571\}$ are neither
optimal nor does the reported PSL of $-113.033724$ correspond to these
coefficients.  The actual PSL corresponding is only $-88.2483$.  The
optimal coefficients were found to be $\{0.00619796,\, 0.10398319,\,
0.49074049\}$ with a corresponding PSL of $-108.2960$, which is better
by $20\,$dB.

The limited computing power of the early 1970s forced Rabiner et
al. to reduce the number of transition parameters whenever possible.
That is why, for example, in the case of bandpass filters, the
transition coefficients on either side were forced to be identical:
``Furthermore, for all cases considered, the bandpass filter samples
were considered to be symmetrical about the center frequency. This
arbitrary constraint is desirable for computational purposes since it
reduces the number of variables by one half. In general, nonsymmetric
transition samples lead to a \textit{somewhat lower} minimax sidelobe,
but this advantage seems canceled out by the increased computational
cost'' \cite[p.~98]{rabiner-gold-mcgonegal-1970} [emphasis added].  We
can only guess that the remark about the PSL being only ``somewhat
lower'' was made based on studying perhaps a small number of BPFs.
Recall that the PSL of the $118$-tap BPF considered earlier was
$-41.2769$ when the transition coefficients were constrained to be
equal.  On the other hand, if we let them to be independent, the
coefficients turn out to be $0.38257443$ and $0.39897929$, resulting
in a lower PSL of $-42.0855$.  The difference is barely $0.8\,$dB.

For the set of $128$-tap BPFs tabulated in Sec.~\ref{sec:fsf_tables},
when making the single transition coefficient on either side
independent, the minimum and maximum improvements were $0.15\,$dB and
$2.14\,$dB, with average being nearly $1\,$dB.  This might lead one to
speculate that the improvements are likely to be roughly in this
range.  This is not so.

\begin{figure}[h!]
\centerline{\includegraphics[width=\columnwidth]{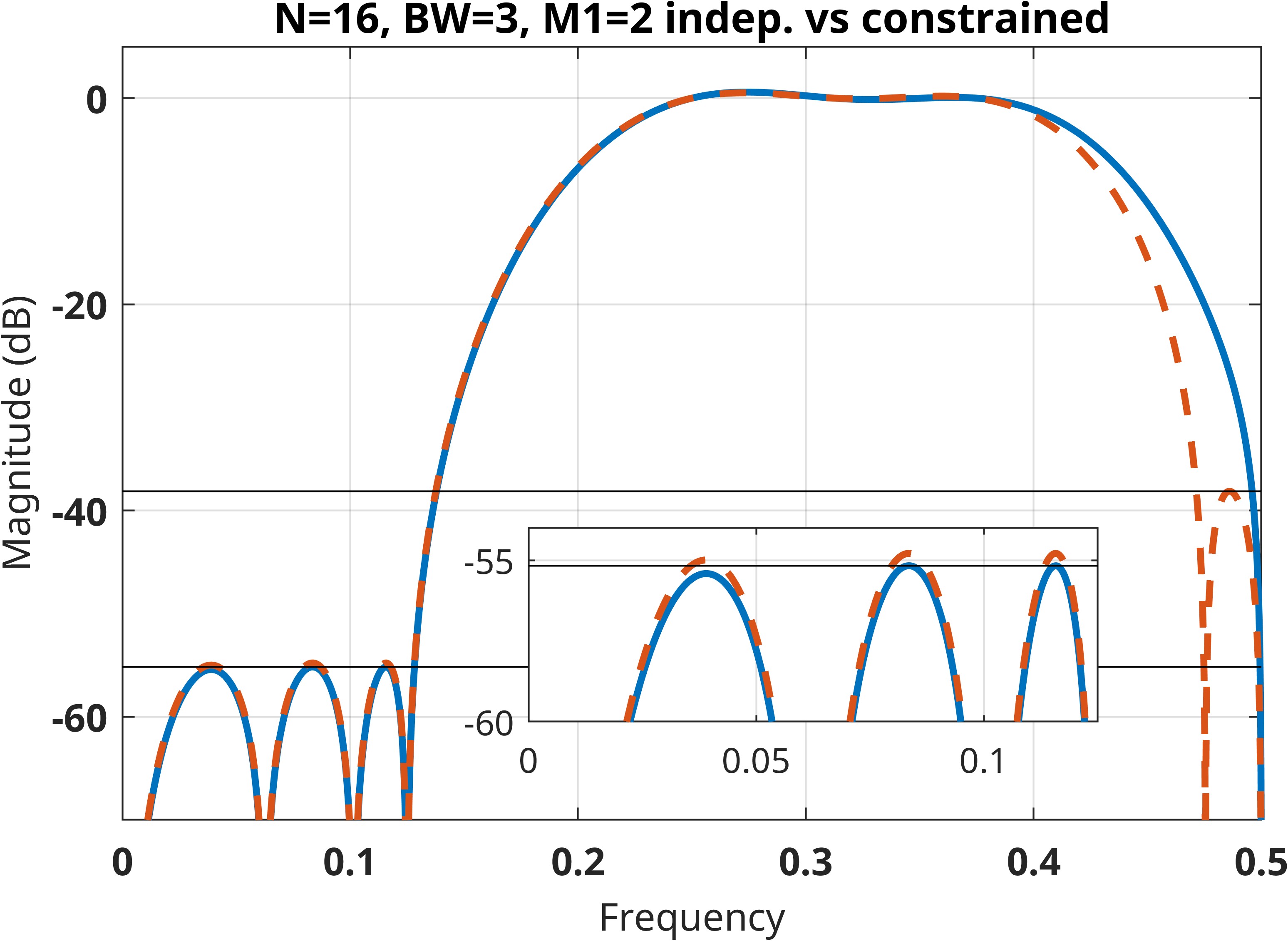}}
\caption{For a BPF with $N=16$, $\mbox{BW}=3$,and $M1=2$, the dashed
  line shows the frequency response of the optimal filter when the
  transition coefficient on one side is constrained to be equal to
  that on the other side.  The optimal coefficient was found to be
  $0.31833319$, having PSL $-38.1280$.  A $17\,$dB improvement is seen
  (solid line) when they are made independent, with optimal values
  $0.31042874$ and $0.45596402$, having PSL $-55.1661$.}
\label{fig:bpf_indep_coeff_1}
\end{figure}

Consider a BPF with $N=16$, $\mbox{BW}=3$, and $M1=2$.  The optimal
value when the coefficients on either side are constrained to be equal
is $0.31833319$, with PSL $-38.1280$.  On the other hand, if we let
the two coefficients to be independent, their values turn out to be
$0.31042874$ and $0.45596402$, but now having PSL $-55.1661$ (see
Fig.~\ref{fig:bpf_indep_coeff_1}), which is a $17\,$dB improvement!
It is interesting to note that the second coefficient's value of
$0.45596402$ is only slightly different from the value of
$0.45593262$ reported in \cite{rabiner-gold-mcgonegal-1970} when the
coefficients are constrained to be equal.

\begin{figure}[t!]
\centerline{\includegraphics[width=\columnwidth]{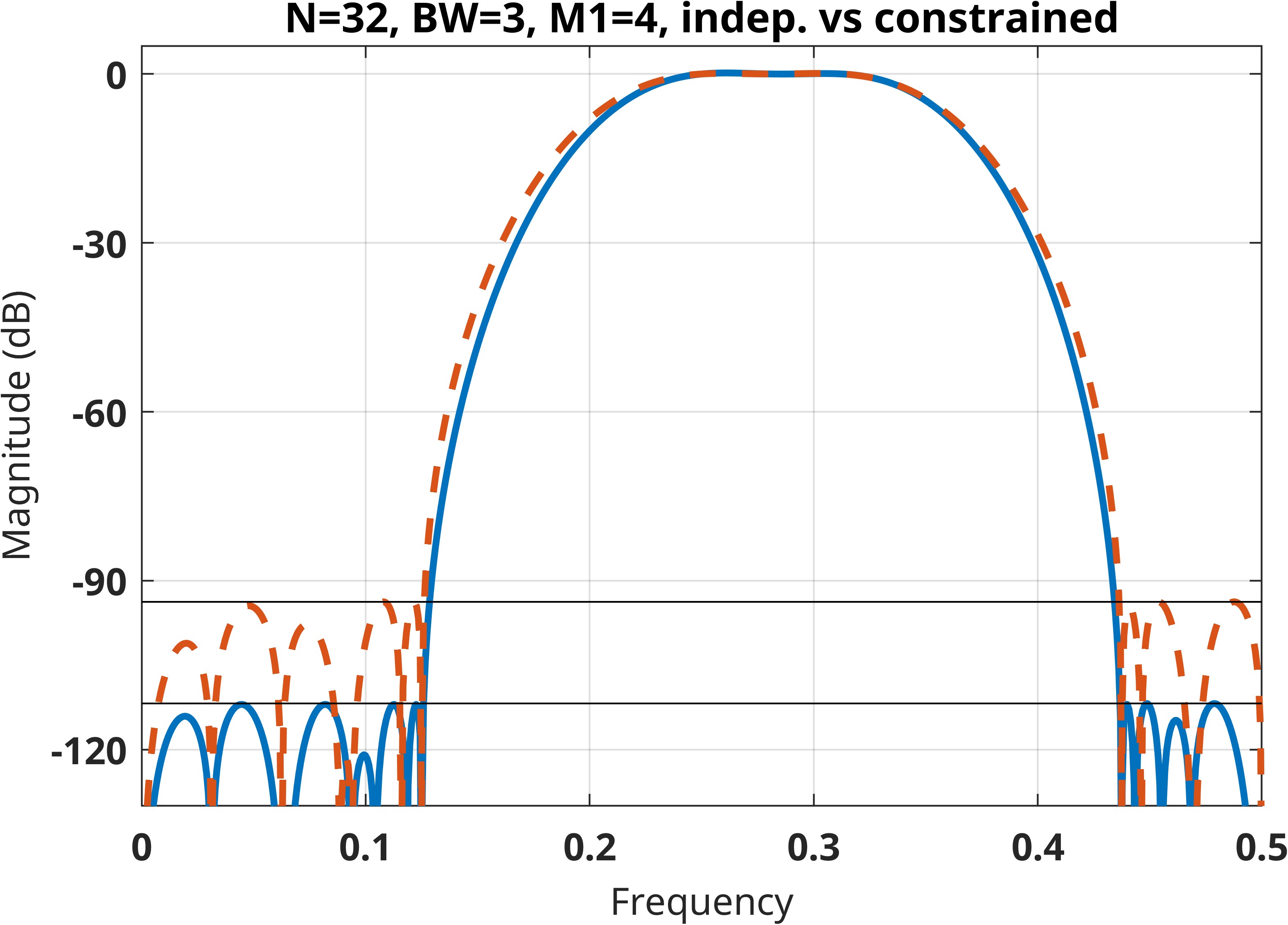}}
\caption{An $18\,$dB improvement in PSL was observed in the case of a
  $32$-tap bandpass filter when the three transition coefficients on
  either side were made independent.  That is, forcing them to be
  equal resulted in a PSL of only $-93.7448$, whereas, making the
  independent reduced it to $-111.7948$.}
\label{fig:bpf_indep_coeff_2}
\end{figure}

An $18\,$dB improvement was observed for $N=32$, $\mbox{BW}=3$, and
$M1=4$ by allowing the three pairs of transition coefficients to be
independent.  When the coefficients were constrained to be equal,
their values were $0.01996720$, $0.22398229$, and $0.70322919$, with a
PSL of $-93.7521$ (the values published in
\cite[p.~99]{rabiner-gold-mcgonegal-1970} are suboptimal).  When they
were allowed to be independent, the estimated values were
$0.00924465$, $0.15712976$, $0.62806540$, $0.68888716$, $0.19070263$,
$0.01184814$, and the PSL was $-111.7948$ (see
Fig.~\ref{fig:bpf_indep_coeff_2}).

A point to note while plotting the magnitude responses of FSFs is that
they should \textit{not} be normalized by their peak passband
values. This is because minimizing the PSL is the main objective, and,
if normalized, the PSL will appear to be lower than what it truly is
(and therefore misleading). Hence, the normalization shown in
\cite{lyons-understanding-dsp-3e}, e.g., Fig.~7-42 (b) and (d), should
be avoided.

From the examples presented so far, it should not be concluded that
there are systematic and significant errors in \textit{all} the
results tabulated in \cite{rabiner-gold-mcgonegal-1970}.  Consider the
LPF with $N=33$, $\mbox{BW}=4$, and one transition coefficient.  The
optimum coefficient is $0.39637169$ with PSL $-42.4538$.  The
transition coefficient's value reported in
\cite{rabiner-gold-mcgonegal-1970} is $0.39641724$.  The PSL
corresponding to this value is $-42.4433$\footnote{The PSL value of
  $-42.45948601$ reported therein is incorrect, especially since it is
  lower than the optimum PSL of $-42.4538$.}, which is within
$0.01\,$dB of our value.  On the other hand, the corresponding values
reported in Lyons \cite{lyons-understanding-dsp-3e} are $0.38433890$
and $-44.4$.  For this transition coefficient, the actual PSL is only
$-40.9768$, which is clearly suboptimal.

A similar match was seen for a $33$-tap LPF with the two transition
coefficients and $\mbox{BW}=9$.  The coefficient values reported in
\cite{rabiner-gold-mcgonegal-1970} are $0.58771575$ and $0.10502930$;
the corresponding PSL is $-67.8157$ (the reported PSL of
$-67.85412312$ is an overestimate by a small margin).  These are close
to the optimum values, viz., $0.58718549$ and $0.10485573$, with PSL
equal to $-68.6612$.  However, the values reported in
\cite{lyons-understanding-dsp-3e} are noticeably different, viz.,
$0.54128598$ and $0.08116809$.  These coefficients were found to be
far from optimal: the corresponding PSL was only $-53.3841$ and not
the published value of $-75.3$.

It should be clear from what we have presented so far that all the
table entries in
\cite{lyons-understanding-dsp-3e,rabiner-gold-mcgonegal-1970} have to
be reexamined.  For example, in
\cite[Table~I]{belorutsky-savinykh-2016} the values published in
\cite{rabiner-gold-mcgonegal-1970} have been used for comparison
by assuming that these values are correct.  In
Section~\ref{sec:fsf_tables} we list the optimal coefficients and
their corresponding PSLs for the LPFs and BPFs considered in these
earlier works.

Coefficient accuracy is, as can be expected, directly influenced by
the fineness of the sampling in the frequency domain that generates
the linear equations.  As a specific example, let us consider $N=16$,
$\mbox{BW}=4$ and one transition coefficient.  Let the density of
samples in the interval $[0,\pi]$ be governed by $\Delta\omega$.
Table~\ref{sampling-density} gives the results for four different
values of $N\cdot\Delta\omega$.  Clearly, the accuracy increases as
$N\cdot\Delta\omega$ becomes smaller.  However, the price paid is the
increased computational burden.  As a compromise, based on this
example, we have chosen $N\cdot\Delta\omega = 0.001$ to obtain values
that are significant up to $8$ digits.
\vspace*{-2mm}
\begin{table}[h!]
  \label{sampling-density}
  \footnotesize
  \centering
  \caption{Effect of density of the frequency samples on the estimated
    transition coefficient for $N=16$, $\mbox{BW}=4$}
  \vspace*{-2mm}
  \begin{tabular}{ccc}
    \hline
    \hline
    $N\cdot\Delta\omega$&Transition Coefficient& PSL (dB) \\
    \hline
    $0.01$    & $0.404742275761249$ & $-41.663251385266385$ \\[1mm]
    $0.001$   & $0.404740970291178$ & $-41.663557574859773$ \\[1mm]
    $0.0001$  & $0.404740968309261$ & $-41.663558039713820$ \\[1mm]
    $0.00001$ & $0.404740968137492$ & $-41.663558080001771$ \\[1mm] \hline
  \end{tabular}
  \vspace*{-2mm}
\end{table}

Depending on the bandwidth, filter length, and number of transition
samples, even a small change in coefficient values can affect the PSL
significantly.  As an example, consider $N=33$, $\mbox{BW}=8$, and
three transition samples.  In Table~\ref{three_trans_coeffs} our
estimates are compared with those given in
\cite{lyons-understanding-dsp-3e,rabiner-gold-mcgonegal-1970}.  Even
though the values appear to be reasonably close, the PSL values are
noticeably different.  In particular, the PSL value of $-92.90$
reported in \cite{lyons-understanding-dsp-3e} for the given
coefficient values is incorrect, the actual value being $-77.6335$.
Clearly, the published values are not optimal.

\begin{table}[H]
  \footnotesize
  \centering
  \caption{An example showing large variations in PSL despite small
    changes in the transition coefficients for $N=33$ and $\mbox{BW}=8$}
  \label{three_trans_coeffs}
  \begin{tabular}{rcccc}
    \hline
    \hline
     & $T_1$&$T_2$&$T_3$ & PSL (dB) \\
    \hline
    Optimal & $0.69966784$ & $0.22290377$ & $0.01921459$ & $-96.1253$ \\[1mm]
    Rabiner & $0.70362590$ & $0.22815933$ & $0.02062988$ & $-92.8212$ \\[1mm]
    Lyons   & $0.70271751$ & $0.22868478$ & $0.02098636$ & \sout{$-92.90$} \\[1mm]\hline
     &  &  & & $-77.6335$
  \end{tabular} 
  \vspace*{-2mm}
\end{table}

In their formulation Rabiner et al. \cite{rabiner-gold-mcgonegal-1970}
classified FSFs as belonging to either Type~I or Type~II.  In our
opinion this nomenclature should have been avoided because Types~I
through IV have been used for the four classes of linear phase filters
\cite{jackson-96,rabiner-gold-75}.  Lyons
\cite{lyons-understanding-dsp-3e} has added significantly to the
confusion by classifying his FSFs as belonging to Types-I through IV
based on the implementation structure.  The statement, ``During this
development, we realized that the even-$N$ Type-I, -II, and -III real
FSFs cannot be used to implement linear-phase highpass filters''
\cite[p.~391]{lyons-understanding-dsp-3e} will undoubtedly cause
confusion because the classification of linear phase filters
themselves into Types~I through IV is standard and well-entrenched in
the literature.  We therefore suggest that these FSF structures be
renamed as Type~A through Type~D to eliminate potential confusion.  In
\cite{lyons-understanding-dsp-3e} Type-IV represents an efficient
real-valued FSF implementation, and further classified as Case~I and
Case~II.  The results presented in Appendix~H
\cite{lyons-understanding-dsp-3e} have been obtained using the Type~IV
implementation.  In our opinion, the terminology for FSFs adopted by
Jackson \cite{jackson-96} is perhaps the best for avoiding any
potential confusion: the filters are classified based on whether they
use the cosine expansion or the sine expansion.  We suggest that it be
used henceforth.

\vspace*{-1mm}
\section{Tables of Corrected Coefficients and PSLs}
\label{sec:fsf_tables}
In Table~\ref{one_trans_coeff_cmp} we give a \textit{comparative
  listing} of coefficients and PSLs for a few LPFs having a single
transition coefficient.  The order of listing is (a) optimum, (b)
Rabiner et al., and (c) Lyons.  Note that, because results for $N=16$
and $\mbox{BW}=5, 6$ have not been reported in
\cite{lyons-understanding-dsp-3e}, there are only two sets of entries
for these BW values.  Also listed are the values published in
\cite{rabiner-gold-mcgonegal-1970,lyons-understanding-dsp-3e}.  The
entries that have been crossed out represent PSL values that have been
reported but are incorrect.  The correct values can be found in the
preceding column.  These entries illustrate the optimality of our
results.
\begin{table}[H]
  \footnotesize
  \centering
  \caption{Optimum coefficients and corresponding PSLs are shown,
    along with published values and corrections to \\ the reported PSLs.}
  \label{one_trans_coeff_cmp}
  \vspace*{-5mm}
\begin{tabularx}{\columnwidth}{YYYY}\\
    \hline
    \hline
    BW & PSL (dB) & & $T_1$ \\\hline
    \multicolumn{3}{c}{$N=16$}   \\\hline
    \multirow{3}{*}{1} & -42.4198 & & 0.43306933 \\
    & -41.2697 & \sout{-39.75363} & 0.42631836 \\
    & -40.1413 & \sout{-44.90}\phantom{000} & 0.41924081 \\\hline
    \multirow{3}{*}{2} & -41.3899 & & 0.41702850 \\
    & -39.7227 & \sout{-37.6135} & 0.40397949 \\
    & -38.1103 & \sout{-45.80}\phantom{~~~} & 0.38969818 \\\hline
    \multirow{3}{*}{3} & -41.2493 & & 0.41001590 \\
    & -39.1661 & \sout{-36.5772} & 0.39454346 \\
    & -36.5836 & \sout{-47.30}\phantom{~~~} & 0.36942214 \\\hline
    \multirow{3}{*}{4} & -41.6636 & & 0.40474097 \\
    & -39.1337 & \sout{-35.8725} & 0.38916626 \\
    & -34.7897 & \sout{-49.60}\phantom{~~~} & 0.34918551 \\\hline
    \multirow{2}{*}{5} & -43.2175 & & 0.39666905 \\
    & -41.1130 & \sout{-35.3170} & 0.38840332 \\\hline
    \multirow{2}{*}{6} & -49.3332 & & 0.37587659\\
    & -40.2540 & \sout{-35.5195} & 0.40155639 \\\hline
    \multicolumn{3}{c}{$N=33$}   \\\hline
    \multirow{3}{*}{1} & -43.0041 & & 0.42980382 \\
    & -42.9621 & \sout{-43.0316} & 0.42994995 \\
    & -42.4198 & \sout{-43.60}\phantom{~~~} & 0.42659097 \\\hline
    \multirow{3}{*}{2} & -42.4130 & & 0.41035321 \\
    & -42.3953 & \sout{42.4253} & 0.41042481 \\
    & -41.5038 & \sout{-43.50}\phantom{~~~} & 0.40433257 \\\hline
    \multirow{3}{*}{3} & -42.4017 & & 0.40135763 \\
    & -42.3879 & \sout{-42.4090} & 0.40141601 \\
    & -41.2145 & \sout{-44.00}\phantom{~~~} & 0.39239983 \\\hline
    \multirow{3}{*}{4} & -42.4538 & & 0.39637170 \\
    & -42.4433 & \sout{-42.4595} & 0.39641724 \\
    & -40.9768 & \sout{-44.40}\phantom{~~~} & 0.38433890 \\\hline
    \multicolumn{3}{c}{$N=128$}   \\\hline
    \multirow{3}{*}{1} & -43.1751 & & 0.42882545 \\
    & -43.1553 & \sout{-43.1530} & 0.42889404 \\
    & -43.1428 & \sout{-43.20}\phantom{~~~} & 0.42864273 \\\hline
  \end{tabularx}
\end{table}

In the following tables, we provide only the optimum transition
coefficient values and their PSLs.

\begin{table}[t!]
  \footnotesize
  \centering
  \caption{LPF, cosine expansion, one and two transition coefficients}
  \label{lpf_cosine_one_and_two_trans_coeff_1}
  \vspace*{-4mm}
  \begin{tabular}{cccccc}\\
    \hline
    \hline
    PSL & $T_1$ & BW & PSL & $T_1$ & $T_2$ \\\hline
    \multicolumn{6}{c}{$N=15$}   \\
    -42.2963 & 0.43370490 & 1 &  -71.8887 & 0.58518349 & 0.09150533 \\
    -41.2619 & 0.41793081 & 2 &  -71.2998 & 0.58546576 & 0.09754605 \\
    -41.2532 & 0.41047294 & 3 &  -73.5765 & 0.57265766 & 0.09175074 \\
    -41.9434 & 0.40402635 & 4 &  -80.0717 & 0.54810045 & 0.07804740 \\
    -44.3219 & 0.39255217 & 5 & -105.5856 & 0.49824241 & 0.05124120 \\
\hline
  \end{tabular}
\end{table}
\afterpage{\clearpage}

\begin{table}[t!]
  \setcounter{table}{3}
  \footnotesize
  \centering
  \caption{LPF, cosine expansion, one and two transition coefficients}
  \label{lpf_cosine_one_and_two_trans_coeff_2}
  \vspace*{-4mm}
  \begin{tabular}{cccccc}\\
    \hline
    \hline
    PSL & $T_1$ & BW & PSL & $T_1$ & $T_2$ \\\hline
    \multicolumn{6}{c}{$N=48$}   \\
    -42.6239 & 0.39972530 &  3 &  -67.6629 & 0.59727667 & 0.10823543 \\
    -42.7607 & 0.39409899 &  4 &  -67.1229 & 0.59727391 & 0.10953933 \\
    -42.8748 & 0.39050686 &  5 &  -66.8027 & 0.59697330 & 0.11014277 \\
    -42.9701 & 0.38804306 &  6 &  -66.5926 & 0.59664297 & 0.11046051 \\
    -43.0426 & 0.38631149 &  7 &  -66.4721 & 0.59623920 & 0.11057077 \\
\hline
  \end{tabular}
\end{table}

\begin{table}[t!]
  \setcounter{table}{3}
  \footnotesize
  \centering
  \caption{LPF, cosine expansion, one and two transition coefficients}
  \label{lpf_cosine_one_and_two_trans_coeff_3}
  \vspace*{-4mm}
  \begin{tabular}{cccccc}\\
    \hline
    \hline
    PSL & $T_1$ & BW & PSL & $T_1$ & $T_2$ \\\hline
    \multicolumn{6}{c}{$N=96$}   \\
    -43.7557 & 0.37724216 & 14 &  -67.2588 & 0.58841865 & 0.10689989 \\
    -43.8285 & 0.37620753 & 17 &  -67.2595 & 0.58777879 & 0.10668569 \\
    -43.8445 & 0.37597462 & 18 &  -67.2551 & 0.58764610 & 0.10664713 \\
    -43.8723 & 0.37553083 & 21 &  -67.2223 & 0.58746696 & 0.10663133 \\
    \multicolumn{6}{c}{$N=125$}   \\
\hline
  \end{tabular}
\end{table}
\afterpage{\clearpage}

\begin{table}[t!]
  \setcounter{table}{3}
  \footnotesize
  \centering
  \caption{LPF, cosine expansion, one and two transition coefficients}
  \label{lpf_cosine_one_and_two_trans_coeff_4}
  \vspace*{-4mm}
  \begin{tabular}{cccccc}\\
    \hline
    \hline
    PSL & $T_1$ & BW & PSL & $T_1$ & $T_2$ \\\hline
    \multicolumn{6}{c}{$N=191$}   \\
    -44.2261 & 0.37174333 & 34 &  -67.9716 & 0.58222388 & 0.10384913 \\
    -44.2542 & 0.37140034 & 41 &  -67.9985 & 0.58188459 & 0.10369859 \\
    -44.2563 & 0.37137303 & 42 &  -67.9992 & 0.58186172 & 0.10368980 \\
    -44.2587 & 0.37131204 & 49 &  -67.9803 & 0.58187600 & 0.10371852 \\
    \multicolumn{6}{c}{$N=192$}   \\
\hline
  \end{tabular}
\end{table}
\afterpage{\clearpage}

\begin{table}[t!]
  \setcounter{table}{3}
  \footnotesize
  \centering
  \caption{LPF, cosine expansion, one and two transition coefficients}
  \label{lpf_cosine_one_and_two_trans_coeff_5}
  \vspace*{-4mm}
  \begin{tabular}{cccccc}\\
    \hline
    \hline
    PSL & $T_1$ & BW & PSL & $T_1$ & $T_2$ \\\hline
    \multicolumn{6}{c}{$N=255$}   \\
    -43.7378 & 0.37802334 & 11 &  -67.5720 & 0.58765486 & 0.10619554 \\
    -44.0002 & 0.37454294 & 17 &  -67.7631 & 0.58482064 & 0.10500078 \\
    -44.0294 & 0.37417868 & 18 &  -67.7919 & 0.58448246 & 0.10484893 \\
    -44.2642 & 0.37137943 & 33 &  -68.0788 & 0.58158144 & 0.10349208 \\
    -44.2724 & 0.37128587 & 34 &  -68.0904 & 0.58147502 & 0.10344091 \\
\hline
  \end{tabular}
  \vspace*{-4mm}
\end{table}
    
\begin{table}[h!]
  \footnotesize
  \centering
  \caption{LPF, cosine expansions, three transition coefficients}
  \label{lpf_cosine_three_trans_coeff_1}
  \vspace*{-4mm}
  \begin{tabular}{ccccc}\\
    \hline
    \hline
    BW & PSL & $T_1$ & $T_2$ & $T_3$ \\\hline
    \multicolumn{5}{c}{$N=15$}   \\
    1 & -105.7660 & 0.65808756 & 0.17066723 & 0.01131839 \\
    2 & -109.4107 & 0.66634773 & 0.18168815 & 0.01170023 \\
    3 & -120.7542 & 0.64705158 & 0.16383941 & 0.00867161 \\
    4 & -160.1316 & 0.60158573 & 0.12372784 & 0.00376140 \\
    \multicolumn{5}{c}{$N=16$}   \\
    1 & -104.2813 & 0.66076794 & 0.17374123 & 0.01187985 \\
    2 & -105.9922 & 0.67271874 & 0.18866558 & 0.01294020 \\
    3 & -115.0748 & 0.65858311 & 0.17537216 & 0.01039969 \\
    4 & -134.4857 & 0.62676044 & 0.14601962 & 0.00625291 \\
    \multicolumn{5}{c}{$N=23$}   \\
    1 & -99.4045 & 0.67060572 & 0.18525334 & 0.01408280 \\
    2 & -96.8845 & 0.69418387 & 0.21314065 & 0.01768526 \\
    3 & -96.9844 & 0.69768235 & 0.21809635 & 0.01826579 \\
    4 & -98.3846 & 0.69377384 & 0.21454407 & 0.01747887 \\
    \multicolumn{5}{c}{$N=24$}   \\
    1 & -99.0821 & 0.67121063 & 0.18598120 & 0.01423144 \\
    \hline
  \end{tabular}
\end{table}
\afterpage{\clearpage}

\begin{table}[h!]
  \setcounter{table}{4}
  \footnotesize
  \centering
  \caption{LPF, cosine expansions, three transition coefficients}
  \label{lpf_cosine_three_trans_coeff_2}
  \vspace*{-4mm}
  \begin{tabular}{ccccc}\\
    \hline
    \hline
    BW & PSL & $T_1$ & $T_2$ & $T_3$ \\\hline
    \multicolumn{5}{c}{$N=24$}   \\
    2 & -96.3422 & 0.69547934 & 0.21468302 & 0.01801263 \\
    3 & -96.4121 & 0.69959240 & 0.22027298 & 0.01870194 \\
    4 & -97.4356 & 0.69683000 & 0.21795421 & 0.01814887 \\
    \multicolumn{5}{c}{$N=32$}   \\
     1 &  -97.7303 & 0.67386181 & 0.18918745 & 0.01489295 \\
\hline
  \end{tabular}
\end{table}
\afterpage{\clearpage}

\begin{table}[t!]
  \setcounter{table}{4}
  \footnotesize
  \centering
  \caption{LPF, cosine expansions, three transition coefficients}
  \label{lpf_cosine_three_trans_coeff_3}
  \vspace*{-4mm}
  \begin{tabular}{ccccc}\\
    \hline
    \hline
    BW & PSL & $T_1$ & $T_2$ & $T_3$ \\\hline
    \multicolumn{5}{c}{$N=96$}   \\
     1 & -98.0013 & 0.67311646 & 0.18832261 & 0.01472738 \\
     2 & -93.8708 & 0.70167067 & 0.22218155 & 0.01964992 \\
     3 & -92.1016 & 0.71182889 & 0.23486317 & 0.02188887 \\
     4 & -91.1478 & 0.71656477 & 0.24101748 & 0.02309218 \\
     8 & -89.7237 & 0.72221922 & 0.24888786 & 0.02481413 \\
\hline
  \end{tabular}
\end{table}

\begin{table}[t!]
  \setcounter{table}{4}
  \footnotesize
  \centering
  \caption{LPF, cosine expansion, three transition coefficients}
  \label{lpf_cosine_three_trans_coeff_4}
  \vspace*{-4mm}
  \begin{tabular}{ccccc}\\
    \hline
    \hline
    BW & PSL & $T_1$ & $T_2$ & $T_3$ \\\hline
    \multicolumn{5}{c}{$N=224$}   \\
     1 & -98.1430 & 0.67279095 & 0.18793677 & 0.01465048 \\
     2 & -94.0467 & 0.70110354 & 0.22151366 & 0.01950969 \\
     3 & -92.3115 & 0.71103432 & 0.23393647 & 0.02169158 \\
     5 & -90.8488 & 0.71787180 & 0.24302568 & 0.02350541 \\
     8 & -90.1121 & 0.72031438 & 0.24673275 & 0.02435580 \\
\hline
  \end{tabular}
  \vspace*{-4mm}
\end{table}

\begin{table}[h!]
  \footnotesize
  \centering
  \caption{LPF, cosine expansions, four transition coefficients}
  \label{lpf_cosine_four_trans_coeff}
  \vspace*{-4mm}
  \begin{tabular}{cccccc}\\
    \hline
    \hline
    BW & PSL & $T_1$ & $T_2$ & $T_3$ & $T_4$ \\\hline
    \multicolumn{6}{c}{$N=16$}   \\
     1 & -146.0821 & 0.70392866 & 0.23366399 & 0.03029531 & 0.00087667\\
    \multicolumn{6}{c}{$N=128$}   \\
    16 & -111.7170 & 0.81122950 & 0.39181283 & 0.08552604 & 0.00519617\\
    \hline
  \end{tabular}
  \vspace*{-4mm}
\end{table}

\begin{table}[b!]
  \footnotesize
  \centering
  \caption{LPF, sine expansion, one and two transition coefficients}
  \label{lpf_sine_one_and_two_trans_coeff_1}
  \vspace*{-4mm}
  \begin{tabular}{cccccc}\\
    \hline
    \hline
    PSL & $T_1$ & BW & PSL & $T_1$ & $T_2$ \\\hline
    \multicolumn{6}{c}{$N=16$}   \\
    -53.9636 & 0.25378977 & 1 &  -85.8184 & 0.38345785 & 0.03967402 \\
    -51.5144 & 0.29471821 & 2 &  -84.5832 & 0.44209941 & 0.04964418 \\
    -51.4189 & 0.30532944 & 3 &  -87.0183 & 0.45185357 & 0.04936662 \\
    -53.4238 & 0.30116938 & 4 &  -98.3962 & 0.43084340 & 0.03953840 \\
    -58.0553 & 0.28356856 & 5 & -128.1412 & 0.38034397 & 0.02312373 \\
    -75.4662 & 0.23382832 & 6 &           &            &            \\
    \hline
  \end{tabular}
\end{table}
\afterpage{\clearpage}

\begin{table}[t!]
  \setcounter{table}{6}
  \footnotesize
  \centering
  \caption{LPF, sine expansion, one and two transition coefficients}
  \label{lpf_sine_one_and_two_trans_coeff_2}
  \vspace*{-4mm}
  \begin{tabular}{cccccc}\\
    \hline
    \hline
    PSL & $T_1$ & BW & PSL & $T_1$ & $T_2$ \\\hline
    \multicolumn{6}{c}{$N=32$}   \\
    -53.1752 & 0.25767056 &  1 &  -82.4539 & 0.39547209 & 0.04415204 \\
    -50.1140 & 0.30310978 &  2 &  -78.2533 & 0.46849535 & 0.06061631 \\
    -48.8917 & 0.32069380 &  3 &  -76.4747 & 0.49704474 & 0.06858542 \\
    -48.2806 & 0.32943588 &  4 &  -75.6783 & 0.51071526 & 0.07267274 \\
    -47.9610 & 0.33420777 &  5 &  -75.3856 & 0.51754997 & 0.07466280 \\
\hline
  \end{tabular}
\end{table}
\afterpage{\clearpage}

\begin{table}[t!]
  \setcounter{table}{6}
  \footnotesize
  \centering
  \caption{LPF, sine expansion, one and two transition coefficients}
  \label{lpf_sine_one_and_two_trans_coeff_3}
  \vspace*{-4mm}
  \begin{tabular}{cccccc}\\
    \hline
    \hline
    PSL & $T_1$ & BW & PSL & $T_1$ & $T_2$ \\\hline
    \multicolumn{6}{c}{$N=256$}   \\
    -45.1338 & 0.36238608 & 33 &  -69.8870 & 0.56812636 & 0.09671288 \\
    -45.1239 & 0.36248904 & 34 &  -69.8702 & 0.56829507 & 0.09679311 \\
    -45.0367 & 0.36340178 & 49 &  -69.7252 & 0.56976108 & 0.09748982 \\
    -45.0338 & 0.36343316 & 50 &  -69.7207 & 0.56980903 & 0.09751231 \\
    -45.0184 & 0.36361749 & 65 &  -69.7057 & 0.57003197 & 0.09760569 \\
    -45.0192 & 0.36361249 & 66 &  -69.7081 & 0.57001604 & 0.09759661 \\
    -45.0598 & 0.36325661 & 81 &  -69.8062 & 0.56923816 & 0.09718744 \\
    -45.0649 & 0.36321044 & 82 &  -69.8182 & 0.56914104 & 0.09713704 \\
    -45.2090 & 0.36191764 &  97 &  -70.1559 & 0.56640952 & 0.09572714 \\
    -45.2258 & 0.36176948 &  98 &  -70.1927 & 0.56610986 & 0.09557391 \\
    -45.8604 & 0.35649731 & 113 &  -71.7497 & 0.55437465 & 0.08957867 \\
    -45.9690 & 0.35565690 & 114 &  -72.0236 & 0.55244168 & 0.08860071 \\
    -47.7521 & 0.34220146 & 121 &  -77.5887 & 0.51868926 & 0.07221001 \\
    -48.4384 & 0.33729826 & 122 &  -80.5087 & 0.50477138 & 0.06581010 \\
    -49.6598 & 0.32955842 & 123 &  -84.9061 & 0.48564315 & 0.05755937 \\
    -52.3950 & 0.31552645 & 124 &  -97.2742 & 0.44980198 & 0.04321282 \\
    -57.5404 & 0.29171189 & 125 & -127.5766 & 0.39063247 & 0.02434591 \\
    -75.2198 & 0.23776312 & 126 &           &            &            \\
    \hline
  \end{tabular}
\end{table}

\begin{table}[b!]
  \footnotesize
  \centering
  \caption{LPF, sine expansion, three transition coefficients}
  \label{lpf_sine_three_trans_coeff_1}
  \vspace*{-4mm}
  \begin{tabular}{ccccc}\\
    \hline
    \hline
    BW & PSL & $T_1$ & $T_2$ & $T_3$ \\\hline
    \multicolumn{5}{c}{$N=16$}   \\
    1 & -123.5411 & 0.46234349 & 0.08532589 & 0.00383979 \\
    2 & -126.8612 & 0.52775336 & 0.10402439 & 0.00445106 \\
    3 & -139.6982 & 0.52842324 & 0.09683625 & 0.00327725 \\
    4 & -184.4643 & 0.48812533 & 0.07000579 & 0.00122350 \\
    \multicolumn{5}{c}{$N=32$}   \\
\hline
  \end{tabular}
\end{table}
\clearpage

\begin{table}[t!]
  \setcounter{table}{7}
  \footnotesize
  \centering
  \caption{LPF, sine expansion, three transition coefficients}
  \label{lpf_sine_three_trans_coeff_2}
  \vspace*{-4mm}
  \begin{tabular}{cccccc}\\
    \hline
    \hline
    BW & PSL & $T_1$ & $T_2$ & $T_3$ \\\hline
    \multicolumn{5}{c}{$N=256$}   \\
     32 &  -92.2882 & 0.70296408 & 0.23038854 & 0.02145714 \\
     48 &  -91.9982 & 0.70497803 & 0.23232644 & 0.02180478 \\
     64 &  -91.9882 & 0.70521938 & 0.23252203 & 0.02183180 \\
     80 &  -92.2325 & 0.70403258 & 0.23126686 & 0.02158219 \\
     96 &  -93.0301 & 0.70006007 & 0.22715283 & 0.02078271 \\
    112 &  -96.8084 & 0.68345550 & 0.21033082 & 0.01761113 \\
    120 & -108.3432 & 0.63851296 & 0.16872662 & 0.01082979 \\
    121 & -113.5569 & 0.62340238 & 0.15584703 & 0.00900611 \\
    122 & -122.2881 & 0.60047029 & 0.13739256 & 0.00666873 \\
    123 & -137.3057 & 0.56676622 & 0.11270268 & 0.00407803 \\
    124 & -183.4675 & 0.50897719 & 0.07640434 & 0.00137999 \\
    \hline
  \end{tabular}
  \vspace*{-4mm}
\end{table}

\begin{table}[h!]
  \footnotesize
  \centering
  \caption{BPF, cosine expansion, up to three transition coefficients}
  \label{bpf_cosine_up_to_three_trans_coeff_1}
  \vspace*{-4mm}
  \begin{tabular}{cccccc}\\
    \hline
    \hline
    BW & M1 & PSL & $T_3$ & $T_2$ & $T_1$ \\\hline
    \multicolumn{6}{c}{$N=16$}   \\
    1 & 2 & -88.6938 &            & 0.05521061 & 0.51538902           \\
    3 & 2 & -38.1280 &            &            & 0.31833319 \\
    \multicolumn{6}{c}{$N=32$}   \\
    1 & 2 &  -99.9654 & 0.01341575 & 0.18239336 & 0.66846507 \\
    1 & 3 &  -97.4502 & 0.01413180 & 0.18589731 & 0.67133344 \\
\hline
  \end{tabular}
\end{table}
\afterpage{\clearpage}

\begin{table}[H]
  \setcounter{table}{8}
  \footnotesize
  \centering
  \caption{BPF, cosine expansion, up to three transition coefficients}
  \label{bpf_cosine_up_to_three_trans_coeff_2}
  \vspace*{-4mm}
  \begin{tabular}{cccccc}\\
    \hline
    \hline
    BW & M1 & PSL & $T_3$ & $T_2$ & $T_1$ \\\hline
    \multicolumn{6}{c}{$N=128$}   \\
    20 & 20 & -45.7751 &            &            & 0.35315763 \\
    26 &  8 & -90.7448 & 0.01996045 & 0.22014430 & 0.69109271 \\
    28 &  8 & -69.0845 &            & 0.09172043 & 0.55645535 \\
    30 &  8 & -45.1061 &            &            & 0.35385726 \\
    31 &  2 & -86.7294 & 0.02488786 & 0.24887129 & 0.72207503 \\
    31 &  8 & -86.3809 & 0.02857646 & 0.26552287 & 0.73602786 \\
    31 & 16 & -85.3475 & 0.03003303 & 0.27133726 & 0.74059324 \\
    31 & 20 & -85.4643 & 0.02892136 & 0.26601945 & 0.73580759 \\
    33 &  2 & -66.0456 &            & 0.11169908 & 0.59724625 \\
    33 &  8 & -62.5410 &            & 0.10906247 & 0.59007419 \\
    33 & 16 & -62.7106 &            & 0.11016850 & 0.59238007 \\
    33 & 20 & -63.1082 &            & 0.10695289 & 0.58753822 \\
    35 &  2 & -40.3677 & \phantom{0.12345678}  & \phantom{0.12345678} & 0.36545360 \\
    35 &  8 & -41.3159 &            &          & 0.37486947 \\
    35 & 16 & -41.0938 &            &          & 0.37995017 \\
    35 & 20 & -40.4066 &            &          & 0.38442987 \\
    \hline
  \end{tabular}
\end{table}
\afterpage{\clearpage}

\balance
\bibliographystyle{IEEEtrans}

\end{document}